\documentclass[12pt, aps,amsmath,showpacs,superscriptaddress]{revtex4-1}
\usepackage{graphicx,dcolumn,subfig,caption,bm,amssymb,amsmath,ulem,indentfirst,amsthm,color}
\usepackage{natbib}
\begin{document}
\title{Short-time particle motion in one and two-dimensional lattices with site disorder}

\author{Bingyu Cui}
\affiliation{Department of Chemistry, University of Pennsylvania, Philadelphia, Pennsylvania 19104,
USA}
\affiliation{School of Chemistry, Tel Aviv University, Tel Aviv 69978, Israel}

\author{Maxim Sukharev}
\affiliation{Department of Physics, Arizona State University, Tempe, Arizona 85287, USA}
\affiliation{College of Integrative Sciences and Arts, Arizona State University, Mesa, Arizona 85201, USA}

\author{Abraham Nitzan}
\email{anitzan@sas.upenn.edu}
\affiliation{Department of Chemistry, University of Pennsylvania, Philadelphia, Pennsylvania 19104,
USA}
\affiliation{School of Chemistry, Tel Aviv University, Tel Aviv 69978, Israel}

\date{\today}

\begin{abstract}
\noindent Like a free particle, the initial growth of a broad (relative to lattice spacing) wavepacket placed on an ordered lattice is slow (zero initial slope) and becomes linear in $t$ at long time. On a disordered lattice, the growth is inhibited at long time (Anderson localization). We consider site disorder with nearest-neighbor hopping on 1- and 2-dimensional systems, and show via numerical simulations supported by the analytical study that the short time growth of the particle distribution is faster on the disordered lattice than on the ordered one. Such faster spread takes place on time and length scale that may be relevant to the exciton motion in disordered systems.
\end{abstract}

\maketitle
\section{Introduction}
Quantum transport in simple dynamic disordered systems has attracted much attention from theorists during the last several decades \cite{Ovchinnikov1974,Madhukar1977,Heinrichs1983,Inaba1981,Heinrichs1983b}.  Under strong disorder, Anderson localization implies that the transport is prohibited beyond a characteristic localization length, with detailed behavior that depends on system dimensionality \cite{Anderson1958,Malyshev2004, Thiessen2017, Kemp2016, Djelti2013, de_Falco2013,Eisfeld2010}. In a 1-dimensional disordered system, Mott and Twose \cite{Mott1961} find that all states are exponentially localized, regardless of the amount of disorder, which is later confirmed and extended to 2-dimensional systems by Abrahams et al. \cite{Abrahams1979}. Such low dimensional disordered systems might range from local site energy disorder in tight-binding models, to those with long-range couplings \cite{Malyshev2003,Kawa2020}. In the case of dynamic disorder that might be induced by the thermal motion of the underlying lattice, short time transport may be faster than in the ordered lattice and becomes diffusive at long time so that the mean square displacement scales as $\langle x^2(t)\rangle\sim t$ when $t\rightarrow\infty$. Closely related are lattice models that describe quantum diffusion on a linear one-band tight binding lattice, with atomic site energies fluctuating in time \cite{Ovchinnikov1974,Madhukar1977,Inaba1981,Girvin1979}.

Generally speaking, disorder is expected to inhibit transport as is most critically realized when localization predominates, while dynamic disorder, including thermal effects, is a source of enhanced transport in such systems. The focus of most work on statically disordered systems is the long time localization issue. Here, we draw attention to another aspect of transport in disordered systems: by combining numerical and analytical studies, we short that on a 1-dimensional disordered lattice the short time spread of an initially prepared carrier wavepacket is faster than the ballstic growth of the same wavepacket on a perfect monoatomic chain. Numerical studies in two dimensions show a similar behavior: at short time (before the localization length is reached) the spread of an initially prepared particle (or exciton) wavepacket is actually enhanced by static disorder.

An important application of these concepts is found in the field of exciton dynamics \cite{Jang2020}. On one hand, exciton transport in static disordered systems is inhibited by localization \cite{Hegarty1985,Barford2013,Singh2017}. On the other hand, it is assisted by exciton-phnonon interaction and becomes diffusive beyond a characteristic coherence length \cite{Barford2013,Singh2017,Qi2021}. Importantly, decay and recombination imply that considerations of these dynamics are relevant only within the finite exciton lifetime that determines also the so called exciton diffusion length, of order $\sim$10-100nm \cite{Menke2014,Abasto2012,Mikhnenko2015,Firdaus2020}. This implies that in such systems the early time dynamics investigated here may be more relevant to the observed dynamics than considerations involving disorder-induced localization. To be specific, we use below the language of free exciton propagation on a lattice of 2-level emitters. Obviously, the same model is relevant for the motion of non-interacting electrons on a disordered lattice of 1-level sites.

The paper is organized as follows: In Section II, we introduce the model and describe numerical simulations that demonstrate this behavior. In particular, we find that the disorder-induced exciton-spread enhancement is a coherent effect that strongly depends on the width of the excitation zone. Excitation spot-size as small as $20$nm can be achieved by near field excitation sources \cite{Novotny1995}, and we find pronounced enhancement for such initial conditions. The effect diminishes for smaller initial excitation spot-sizes and disappears when the initial state is a single excited molecule. In Section III, we confirm the numerical observation by providing an analytical derivation of the short-time behavior of wavepacket width. Section IV concludes. 
%To be specific, we use below the language of free exciton propagation on a (1-d) lattice of two-level emitters. Obviously, the same model is relevant for the motion of non-interacting electrons on a disordered lattice of 1-level sites.

\section{Numerical simulations}
\subsection{Model and simulation procedure}
We consider a linear chain of 2-level emitters with nearest-neighbor coupling $J$. The Hamiltonian is 
\begin{equation}
    \hat{H}=\sum_n\epsilon_n\hat{c}_n^\dagger\hat{c}_n+J\sum_n(\hat{c}_n^\dagger\hat{c}_{n+1}+\hat{c}_n\hat{c}_{n+1}^\dagger)
    \label{eq:Hamiltonian}
\end{equation}
where $\hat{c}_n^\dagger$ and $\hat{c}_n$ respectively create and destroy an excitation on site $n$, and the coupling $J$ moves it between nearest-neighbor sites.
%\textcolor{blue}{Since the typical bandwidth of organic semiconductors is 1eV \cite{?}, in numerical simulations we take the coupling strength to be $0.5$eV.}
The site energies are sampled from a Gaussian distribution with $\langle \epsilon_n\rangle_E=0$ and $\langle\epsilon_{n}\epsilon_{n'}\rangle_E=\langle\epsilon_n^2\rangle_E\delta_{nn'}\equiv\sigma^2\delta_{nn'}$. Here, $\langle...\rangle_E$ denotes the ensemble average, while $\langle...\rangle$ is used below for the quantum mechanical expectation value.
%\textcolor{blue}{We take a range of disorder parameters, $0.1J\leq \sigma\leq J$ in simulations.} 
The chain is taken long enough so that boundary effects are not relevant for the simulated time and length scales. In the reported simulations we have used emitter chains of $2.5\times10^4$ sites, and have ascertained that further increase of the chain length did not affect the computed dynamics. The initial state was taken to be a Gaussian wavepacket with width $D$
\begin{equation}
\Psi(x,t=0)=\frac{\sum_ne^{-\frac{(na)^2}{D^2}}\phi(x-na)}{\sqrt{\sum_ne^{-2\left(\frac{na}{D}\right)^2}}},
\label{eq:psi0}
\end{equation}
where $\phi(x)$ is the orbital wavefunction at position $x$ and $a$ is the lattice spacing and the site wavefunctions $\phi(x-na)$ are assumed to localized at site $x=na$ such that $\langle\phi(x-na)|f(\hat{x})|\phi(x-ma)\rangle=f(x-na)\delta_{nm}$ for an arbitrary function of position $f(x)$ and $\delta_{nm}$ is the Kronecker delta. In simulations reported below the initial values of the Gaussian width were taken to be $D=a,5a,20a$, which corresponds to an initial wavepacket with $\langle x^2(t=0)\rangle=\sum_n(na)^2\exp(-n^2a^2/D^2)/\sum_n\exp(-n^2a^2/D^2)$. 
%\textcolor{blue}{Note that in standard excitations, the spot-size is of the order of radiation wavelength, but to experimentally observed exciton diffusion, one needs to apply near-field excitation with spot-size smaller than the exciton diffusion length ($\sim$100nm or less, see, e.g., Ref. \cite{Sajjad2020}). The largest initial wavepacket width that we have used is 20nm, of the order achievable in such nearfield excitations \cite{Novotny1995}). We have also examined the effect of choosing smaller initial spot-sizes, and as will be shown later, the effect we are describing becomes more pronounced when the initial spot-size is larger.}
The width at time $t$ is calculated as the square root of $\langle \delta x^2(t)\rangle\equiv\langle x^2(t)\rangle-\langle x(t)\rangle^2$, where for any operator $\hat{A}$,
\begin{align}
\langle A(t)\rangle=\langle\Psi(x,t=0)|e^{i\hat{H}t/\hbar}\hat{A}e^{-i\hat{H}t/\hbar}|\Psi(x,t=0)\rangle.
\label{eq:TBG}
\end{align}
This calculation was repeated over many realizations of the disorder lattice and the final result was obtained as an ensemble average over the disorder. The time evolution was calculated by diagonalizing the Hamiltonian.

We consider multiple cases: (1) an ordered system with $\epsilon_n=0$ for all $n$; (2) a system with static disorder characterized by a Gaussian random distribution of site energies with $\langle\epsilon_n\rangle_E=0$ and $\sigma=\langle\epsilon_n^2\rangle^{1/2}_E$ ranging between $0.01J$ and $0.5J$; (3) for completeness we also show results for a dynamic disorder model where values of the site energies were resampled at time intervals $\tau$ that in turn are sampled (unless otherwise stated) from a Poisson distribution characterized by an average renewal time $\langle \tau\rangle$. In all simulations, we calculated the width of the wavepacket as a function of time, averaged over trajectories. We have found that averaging from more than 64 trajectories for nearly all simulation parameters does not noticeably change our results.

\subsection{Numerical results}
In an ordered lattice, the time evolution of the root mean square displacement (RMSD) from the origin is similar to the free particle behavior
\begin{align}
    &\sqrt{\langle \delta x^2\rangle}=\frac{1}{2}\sqrt{D^2+\frac{\hbar^2t^2}{4m^2D^2}}\notag\\
    \rightarrow&\frac{\hbar t}{4mD} \quad \text{as}\quad t\rightarrow\infty,
\end{align}
where the particle mass $m$ is related to the coupling $J$ of Eq. \eqref{eq:Hamiltonian} and the lattice constant $a$ by 
\begin{equation}
    J=\frac{\hbar^2}{2ma^2}.
\end{equation}
Following an initial time of order $2mD^2/\hbar$, in which the wavepacket width increases only slowly, the expansion peaks up and becomes ballistic-like, $\sqrt{\langle \delta x^2\rangle}\sim t$ at long time. The initial incubation period may be discussed in terms of destructive interference between quantum trajectories originating from different sites.

\begin{figure}[!tp]
\subfloat[][]{
\begin{minipage}[t]{0.5\textwidth}
\flushleft
\includegraphics[width=0.8\textwidth]{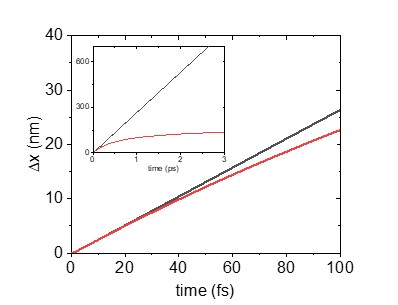}
\end{minipage}
}
\\
\subfloat[][]{
\begin{minipage}[t]{0.5\textwidth}
\flushleft
\includegraphics[width=0.8\textwidth]{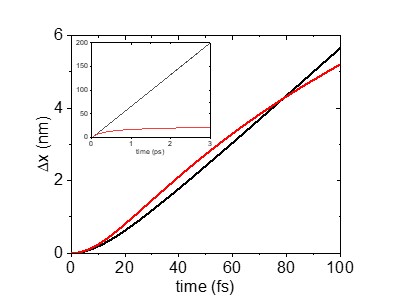}
\end{minipage}
}
\\
\subfloat[][]{
\begin{minipage}[t]{0.5\textwidth}
\flushleft
\includegraphics[width=0.8\textwidth]{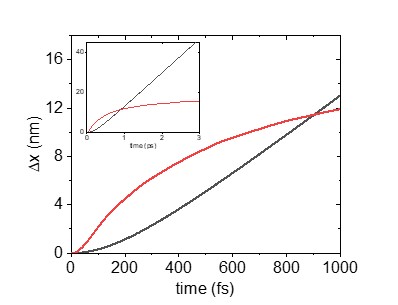}
\end{minipage}
}
\caption{The root mean square displacement Eq. \eqref{eq:MSD} calculated from numerical simulations, shows for ordered (black lines) and static disordered (red lines) cases. The hopping parameter is $J=0.5$eV and the disorder parameter is $\sigma=0.05eV$. The widths of the initial wavepacket are (a) $D=1$nm, (b) $5$nm and (c) $20$nm. Inset panels are dynamics at longer time.}
\label{fig:1}
\end{figure}

In simulations described below, unless otherwise noted, the intersite coupling was taken $0.5$eV which corresponds to a bandwidth of $1$eV in one dimension. The lattice spacing was $a=1$nm and the initial excitation spot-size was taken $20$nm or smaller. Excitation spot-sizes as small as $\sim$20nm \cite{Novotny1995} can be achieved using near-field excitation sources and we have simulated also processes with smaller initial widths as a way to support the proposed origin of the observed behavior.

Figure \ref{fig:1} shows our results for the increase in the averaged width, 
\begin{equation}
\Delta x\equiv\sqrt{\langle \delta x^2(t)\rangle}-\sqrt{\langle \delta x^2(t=0)\rangle},
\label{eq:MSD}
\end{equation}
for several choices of initial wavepacket width $D$, with different panels displaying results for $D/a=1,5,20$. If we accept the picture according to which the slow initial spread reflects destructive interference, this interference appears to erode upon the introduction of site disorder, leading to a significant increase of the spreading rate in this regime. Indeed, when the initial wavepacket width is $20$nm, the expansion rate in the disordered systems exceeds that in the ordered lattice until an excess spread of order $10$nm, which is of the order of diffusion lengths of excitons in bulk heterojunction photovoltaic cells. This is a short time effect: in the present 1-dimensional site-disordered model with nearest-neighbor coupling, all wavefunctions are localized and expansion eventually stops as seen in the insets. 

\begin{figure}[!tp]
\includegraphics[width=0.8\textwidth]{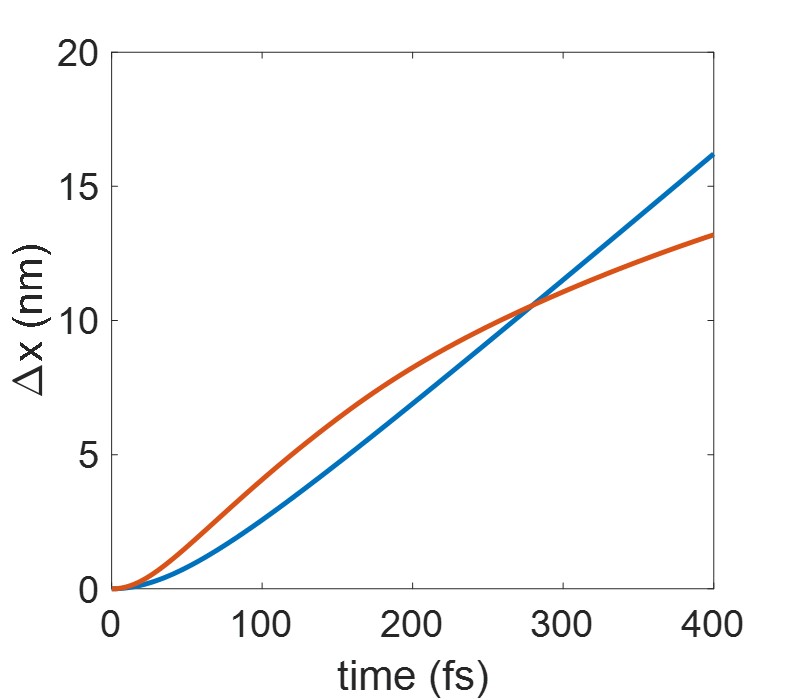}
\caption{The root mean square displacement Eq. \eqref{eq:MSD} calculated from numerical simulations in 2-dimension, shows for ordered (blue line) and static disordered (red line) cases. The hopping parameter is $J=0.5$eV and the disorder parameter is $\sigma=0.1eV$. The width of the initial wavepacket is $D=10$nm.}
\label{fig:2}
\end{figure}

Our interpretation for the observed effect should not be dimensionality dependent. Indeed, Fig. \ref{fig:2} shows a similar effect in a 2-dimensional calculation. Also, while the Gaussian form of the initial exciton wavepacket is a natural choice for this study, we show (see Fig. S1 in the Supplementary Information (SI) - an initial $p$-like state) that the observed effect does not depend on this choice. These observations suggest that the static disorder can cause enhancement of exciton diffusion also in realistic systems, and motivates future studies in this direction.

A direct observation of the predicted short time behavior would require excitation of a small spot-size and zero linear momentum in the observed direction (as may be achieved by exciting a surface exciton using incident field normal to the surface). Nevertheless, we have also studied the time evolution of an initially prepared exciton wavepacket with a finite linear momentum, and Fig. S2 in the SI shows the results of such a study. The effect of the initial linear momentum on the spread of the wavepacket appears to be minimal.

\begin{figure}[!tp]
\includegraphics[width=0.8\textwidth]{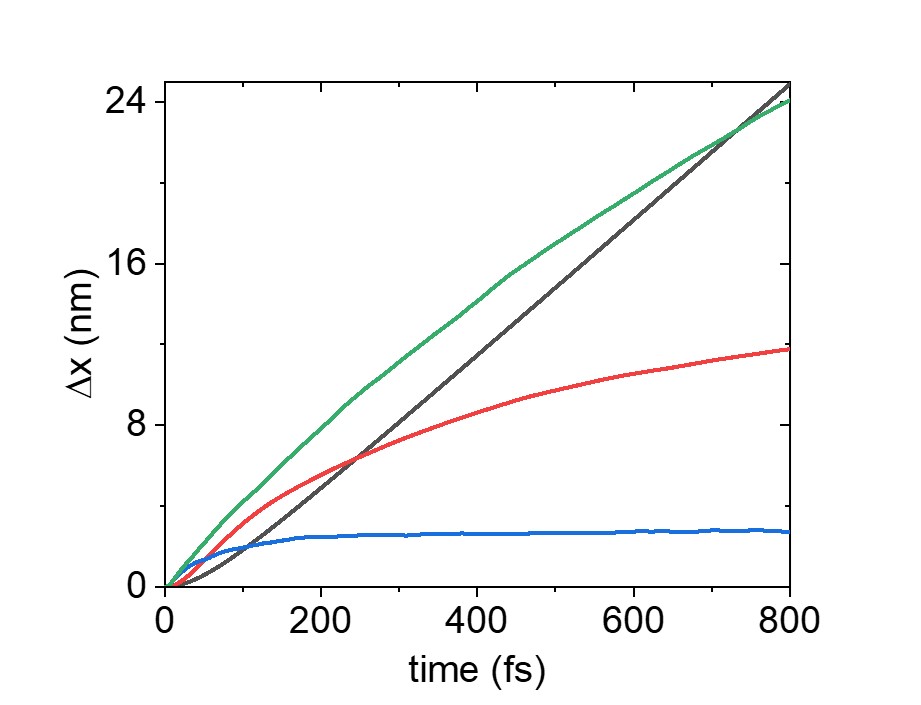}
\caption{The root mean square displacement Eq. \eqref{eq:MSD} calculated from numerical simulations, shows for ordered (black line), static and dynamic disordered cases. The hopping parameter is $J=0.5$eV and the disorder parameters are $\sigma=0.05eV$ (red line) and $\sigma=0.2eV$ (blue line) for static disorder and $\sigma=0.2eV$ for dynamic disorder (green line) in which random renewal kicks are performed at time intervals $\tau$ sampled from a Poisson distribution with $\langle\tau\rangle=52$fs. The width of the initial wavepacket is $D=10$nm. }
\label{fig:3}
\end{figure}

As detailed in the introduction, dynamic disorder has long been connected with acceleration of transport in disordered systems. Recent work on exciton transport has similarly discussed phonon-assisted exciton transport \cite{Jang2020,Barford2013,Troisi2003,Kilgour2015,Mejia2022}. Fig. \ref{fig:3} shows the effect of static and dynamic disorder on carrier mobility in comparison with the underlying ordered lattice. Both static and dynamic disorder are seen to accelerate the expansion rate of an initially formed wavepacket relative to the disordered system, however, the superdiffusion dynamics on the ordered lattice takes over at long time. Expansion under dynamic disorder remains faster than on the ordered lattice for considerably longer time than that under static order, but given that moving carriers are subjected to competing short time processes (emission and charge separation at nearby interfaces for excitons, recombination and absorption at surfaces for electrons), the very short time dynamics where static disorder also has a significant effect is relevant to the operation of many such systems.

Figure \ref{fig:3} provides another interesting observation: acceleration of the wavepacket expansion is more efficient for smaller amplitude of the disorder. A possible explanation is that static disorder has two effects on short-time quantum transport: (a) destroying destructive interference that otherwise inhibits wavepacket propagation and (b) reducing the effect of coherent transport. The observation that the effect of smaller disorder amplitude persists longer than that of the larger one implies that removing interference between quantum trajectories initiated on different lattice sites is the more important short time effect at least for our present choice of parameters.

\begin{figure}[!tp]
\includegraphics[width=0.8\textwidth]{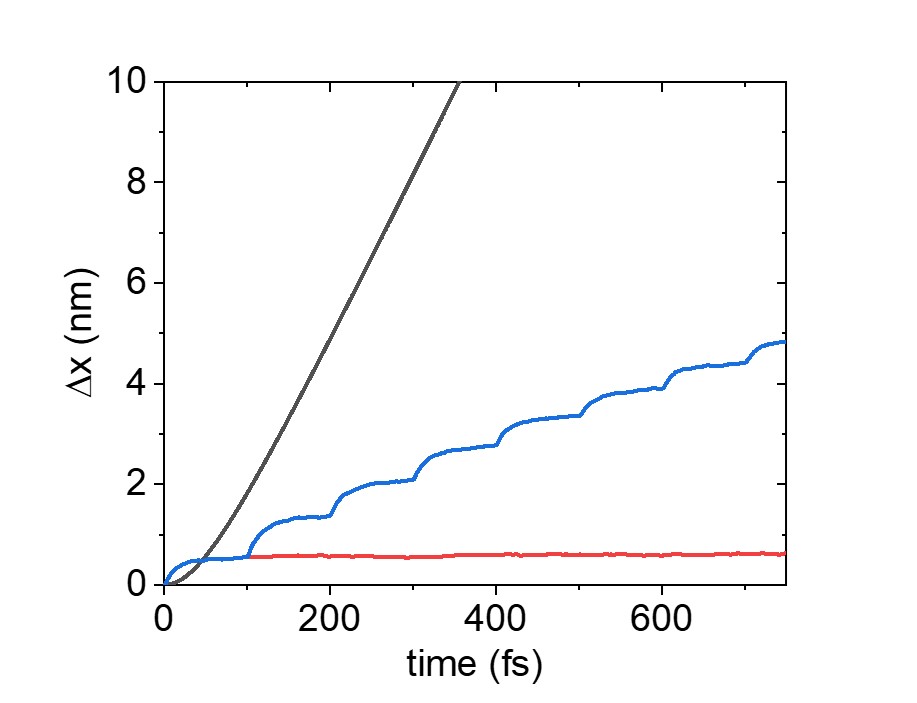}
\caption{The root mean square displacement Eq. \eqref{eq:MSD} calculated from numerical simulations, shows for ordered (black line), static disordered (red line) and dynamic disordered (blue line, disorder is updated at every $\tau=100$fs) cases. The hopping parameter is $J=0.5$eV and the disorder parameter is $\sigma=0.5eV$. The width of the initial wavepacket is $D=10$nm. }
\label{fig:4}
\end{figure}

Finally, we point out the effect of dynamic disorder has its root in the properties of the underlying static disorder. This is seen in Fig. \ref{fig:4} that shows the wavepacket expansion process in a system where dynamic disorder is made by a sequence of disorder updates, made at constant time intervals $\tau$, at which the site energies are resampled from their distribution. Each such update is seen to be followed by enhanced expansion that subsides as the wavepacket explores its new localization region. Together these updates lead to a long-time diffusive expansion that reflects the series of transiently accelerated expansions that follow each update.

\section{Analytical evaluation}
Here we attempt to rationalize the main observations made above that the speed of an excitonic wavepacket is initially accelerated by static disorder, by looking at the short time evolution under the Hamiltonian \eqref{eq:Hamiltonian}. Our goal is to calculate the evolution of the mean size $\Delta x(t)$ of an initially prepared wavepacket, for a 1-dimensional site-disorder model. In what follows we describe a short time approximation for this evolution which is able to describe its initial trend.

We start, following Ref. \cite{Madhukar1977}, with a more general Hamiltonian given in the site representation by
\begin{align}
    H&=\frac{1}{2}\sum_{m,n}\alpha_{mn}\{|m\rangle\langle n|+|n\rangle\langle m|\}\notag\\
    &+\frac{1}{2}\sum_{m,n}\beta_{mn}\{|m\rangle\langle n|+|n\rangle\langle m|\},   
    \label{eq:Hrandom}
\end{align}
where $n$ and $m$ denote sites on a 1-dimensional periodic lattice and $\alpha_{mm}$ and $\beta_{mn}$ denote the deterministic and random parts, respectively, of the Hamiltonian matrix. Specifically, we assume that Eq. \eqref{eq:Hrandom} represents an ensemble of identical tight binding systems, each of which is characterized by the tight binding parameter $J$ so that
\begin{equation}
    \alpha_{mn}=J\delta_{|m-n|,1}
\end{equation}
and by a particular realization of the parameters $\beta_{mn}$, which we take to be Gaussian random variables specified by the ensemble average $\langle\beta_{mn}\rangle_E=0$ and
\begin{equation}
\langle\beta_{mn}\beta_{m'n'}\rangle_E=g(m-n)(\delta_{mm'}\delta_{nn'}+\delta_{mn'}\delta_{m'n}-\delta_{mn}\delta_{m'n'}\delta_{nn'}).
\label{eq:avebeta}
\end{equation} 
Here, $g(m-n)$ measures the strength of the disorder. For thermally induced disorder (i.e., phonons) $g$ reflects the carrier-phonon coupling strength, and generally depends on temperature. In particular, we will focus on the case of site-diagonal; disorder described by $g(m-n)=g(0)\delta_{mn}$. 

In what follows, we follow the approach of Ref. 
\cite{Madhukar1977,Singh2005}, adapting it for the short time dynamics under static disorder. The density matrix satisfies the quantum Liouville equation
\begin{equation}
\frac{\partial  \rho}{\partial t}=-\frac{i}{\hbar}[H,\rho],
\end{equation}
which corresponds to
\begin{align}
\frac{\partial \rho_{l,r}}{\partial t}&=-\frac{i}{\hbar}J(\rho_{l+1,r}+\rho_{l-1,r}-\rho_{l,r+1}-\rho_{l,r-1})\notag\\
&-\frac{i}{2\hbar}\sum_n[(\beta_{ln}+\beta_{nl})\rho_{n,r}-(\beta_{nr}+\beta_{rn})\rho_{l,n}].
\end{align}
For convenience we set the lattice spacing to $a=1$. Taking the (spatial) Fourier transformation, $\tilde{f}(k_1,k_2)=\sum_{lr}e^{-ik_1l+ik_2r}f_{l,r}$ on both sides, as well as the ensemble average over the distribution of the $\beta$ parameters, we obtain

\begin{align}
\frac{\partial\langle \tilde{\rho}(k_1,k_2;t)\rangle_E}{\partial t}&=-\frac{2Ji}{\hbar}(\cos(k_1)-\cos(k_2))\langle\tilde{\rho}(k_1,k_2;t)\rangle_E\notag\\
&-\frac{i}{2\pi\hbar}\int_{-\pi}^\pi\int_{-\pi}^\pi dqdq'\langle[\tilde{\beta}(k_1,q)\delta(q'-k_2)-\tilde{\beta}(q',k_2)\delta(q-k_1)]\tilde{\rho}(q,q';t)\rangle_E.
\label{eq:rhoSF}
\end{align}

\begin{figure}[!tp]
\includegraphics[width=0.8\textwidth]{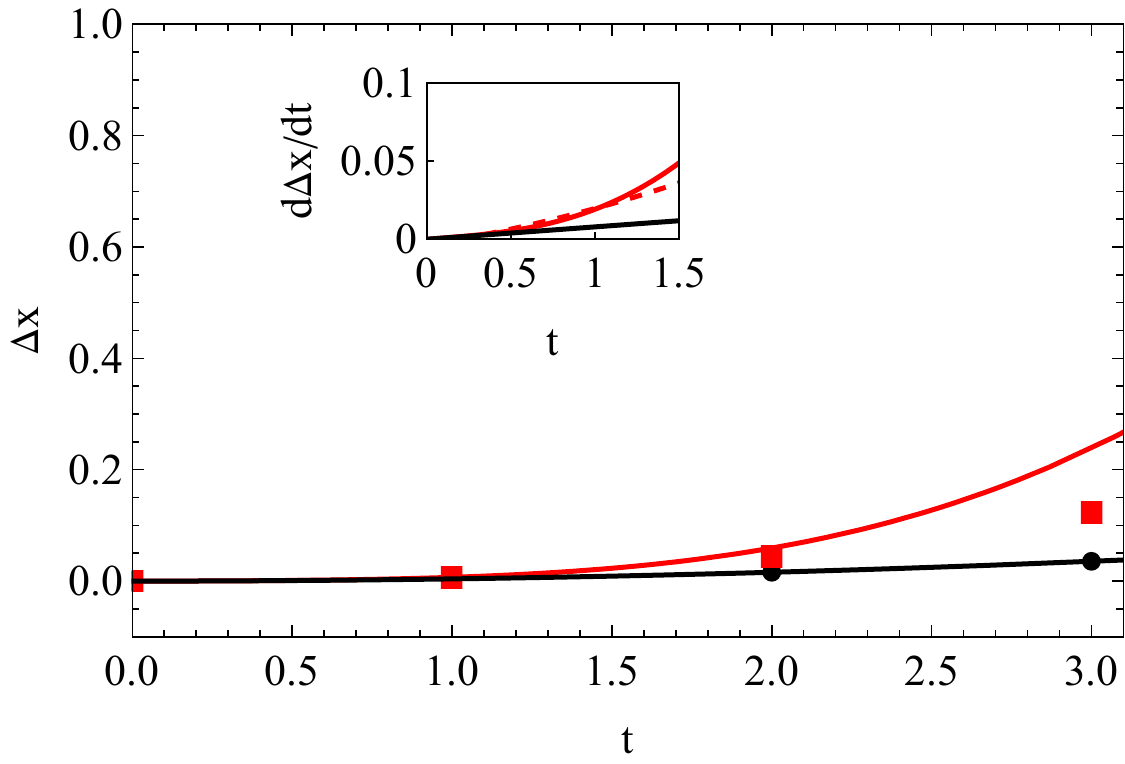}
\caption{The time evolution of the spread $\Delta x$ of the exciton wavepacket calculated from Eq. \eqref{eq:msdG} (solid lines), and from our numerical simulation (circles and squares). The initial exciton width is $10$. Results of the ordered lattice ($g(0)=0$) are shown in solid black line and circles, while those corresponding to the disordered case ($g(0)=0.09J^2$) are displayed in solid red line and red squares. The inset shows the time derivative of change in RMSD, where the red dashed line is the result of interpolation of numerical dots calculated for the disordered system. The simulation cell contains $N=501$ lattice points. An average of $n=60$ realizations is taken and the estimated error in the numerical collection of the disordered system is smaller ($<10\%$) than the size of point.  All number are in dimensional units defined in terms of lattice spacing $a(=1)$ and the nearest neighbor coupling energy $J(=1$) so that the time unit is $\hbar/J$ (for the choice $J=0.5$eV, a unit of time is $\sim$1.25fs and we take $\hbar=1$).}
\label{fig:5}
\end{figure}

The evaluation of a short time solution of Eq. \eqref{eq:rhoSF} is described in Sec. II of the SI, where details on the way the short time assumption is implemented are provided. This calculation leads, for site diagonal disorder, to the Laplace transform $\int_0^\infty dt e^{-st}\langle x^2(t)\rangle$ of the RMSD in the form (see Sec. II in the SI for more details)
\begin{equation}
\langle x^2(s)\rangle=-\left[\frac{\partial^2\tilde{\chi}(u;s)}{\partial u^2}\right]_{u=0},
\label{eq:msdG}
\end{equation}
where 
\begin{equation}
\tilde{\chi}(u;s)\equiv\frac{1}{2\pi}\int_{-\pi}^\pi\langle \hat{R}(p,u;s)\rangle_E dp=\frac{1}{2\pi}\int_{-\pi}^\pi\langle\hat{R}(q+p,u;s)\rangle_E dp,
\end{equation}
and $\hat{R}(p,u;s)\equiv \hat{\tilde{\rho}}(k_1,k_2;s)$ with $k_1=p+u/2;k_2=p-u/2$.
The function $\tilde{\chi}(u;s)$ is found (Sec. II in the SI) to be given by
\begin{align}
\tilde{\chi}(u;s)&=\frac{I_1}{1-2g(0)I_2/\hbar^2},
\label{eq:chi}
\end{align}
where $I_1$ and $I_2$ are given by
\begin{subequations}
\begin{align}
\label{eq:I1}
I_1(s)=&\frac{1}{2\pi}\int_{-\pi}^\pi\frac{R(p,u;t=0) dp}{s-i4J\sin(p)\sin\left(\frac{u}{2}\right)/\hbar+2g(0)/(s\hbar^2)},\\
I_2(s)=&\frac{1}{2\pi}\int_{-\pi}^\pi\frac{dp}{s^2-i4sJ\sin(p)\sin\left(\frac{u}{2}\right)/\hbar+2g(0)/\hbar^2}.
\label{eq:I2}
\end{align}
\label{eq:I}
\end{subequations}
In Eq. \eqref{eq:I}, the form $R(p,u;t=0)$ is obtained from the initial wavepacket, Eq. \eqref{eq:psi0}, and is given by
\begin{equation}
    R(p,u;t=0)=\sqrt{2\pi D^2}e^{-\frac{D^2}{4}\left(2p^2+\frac{u^2}{2}\right)}.
\end{equation}

Finally, $\langle x^2(t)\rangle$ is calculated as the inverse transform of $\langle x^2(s)\rangle$. Figure \ref{fig:5} compares results obtained from this procedure to those calculated from the numerical simulation. While the agreement between these results deteriorates as $t$ increases, the analytical result clearly shows a faster increase in the RMSD for the disordered case in comparison with the ordered system. Note that our approximation (using Eq. (S10) instead of Eq. (S9) in the SI) is rigorously valid only for time shorter than our time unit $\hbar/J$ (we disregard oscillatory terms that appear on a longer timescale) as indeed seen in the inset of Fig. \ref{fig:5}. At longer times, the analytically calculated spread overestimates the simulation results. This behavior may be viewed as consistent with our assertion that the disorder induced spread enhancement is associated with the erasure of destructive interference, provided that this interference is manifested through the aforementioned oscillations.

\section{Conclusion}
Using 1- and 2-dimensional site disorder models, we have found that for an initial exciton (or particle) wavepacket whose width encompasses several sites, the initial spread is accelerated by static disorder. Such disorder affects the time evolution in two ways: First, it disrupts the destructive interference between waves emanating from different sites (hence the initial speed acceleration), second it inhibits later coherent evolution (causing later localization). For the broad enough initial wavepacket (as may be formed by optical excitations) the time and length scales of the accelerated speed may be of the order of the excitonic lifetimes and diffusion length. Extending the present findings to 3-dimensional systems will be a subject to future study.

%We have shown in this paper that the motion of a particle on a disordered solid exhibits features more than localizing in the long time purpose. The Hamiltonian of the particle is divided into two parts, one representing the motion of the atom on the rigid surface, the other representing the random fluctuations that modify the site energies. The mean square displacement is proportional to $t^2$ in the former case. If these fluctuations are static Gaussian process and a constant nearest neighbor interaction is assumed, the short-time evolution of mean square displacement of the particle can be calculated, which promotes the ballistic motion of the particle. Such disorder is able to quench superradiance, which would also shed light in the research of nanoscopic systems in quantum coherent regime \cite{Celardo2013,Celardo2014}.

%Arguments presented here are only limited to static disorder. It would be of interest to see if systems undergoing different dynamic disorders would exhibit similar behaviors. Also, the generalization of this work to higher dimensions is another a possible future direction.

\section*{Supplementary Material}
See supplementary material for examples of the time-dependent spread of an initial wavepacket carrying non-zero momentum, the detailed derivation of the analytical results discussed in Sec. III and the demonstration that our analytical approach reproduces the exact dynamics in the ordered lattice case.

\section*{Acknowledgements}
This material is based upon work supported by the U.S. National Science Foundation under Grant CHE1953701. M.S. is supported by the Air Force Office of Scientific Research under Grant No. FA9550-22-1-0175. We thank Joe Subotnik for many useful discussions.

\section*{Author Declarations}
\subsection*{Conflict of interest}
The authors have no conflicts to disclose. 

\section*{Data availability}
The data that support the findings of this study are available within the article [and its supplementary material].

\bibliography{Anderson}

\end{document}

\title{Supplementary Information for "Short-time particle motion in one and two-dimensional lattices with site disorder"}

\author{Bingyu Cui}
\affiliation{Department of Chemistry, University of Pennsylvania, Philadelphia, Pennsylvania 19104,
USA}
\affiliation{School of Chemistry, Tel Aviv University, Tel Aviv 69978, Israel}

\author{Maxim Sukharev}
\affiliation{Department of Physics, Arizona State University, Tempe, Arizona 85287, USA}
\affiliation{College of Integrative Sciences and Arts, Arizona State University, Mesa, Arizona 85201, USA}

\author{Abraham Nitzan}
\email{anitzan@sas.upenn.edu}
\affiliation{Department of Chemistry, University of Pennsylvania, Philadelphia, Pennsylvania 19104,
USA}
\affiliation{School of Chemistry, Tel Aviv University, Tel Aviv 69978, Israel}

\date{\today}
\maketitle

\section{Dependence of the wavepacket spread on initial shape and linear momentum}

\begin{figure}[!htp]
\includegraphics[width=0.8\textwidth]{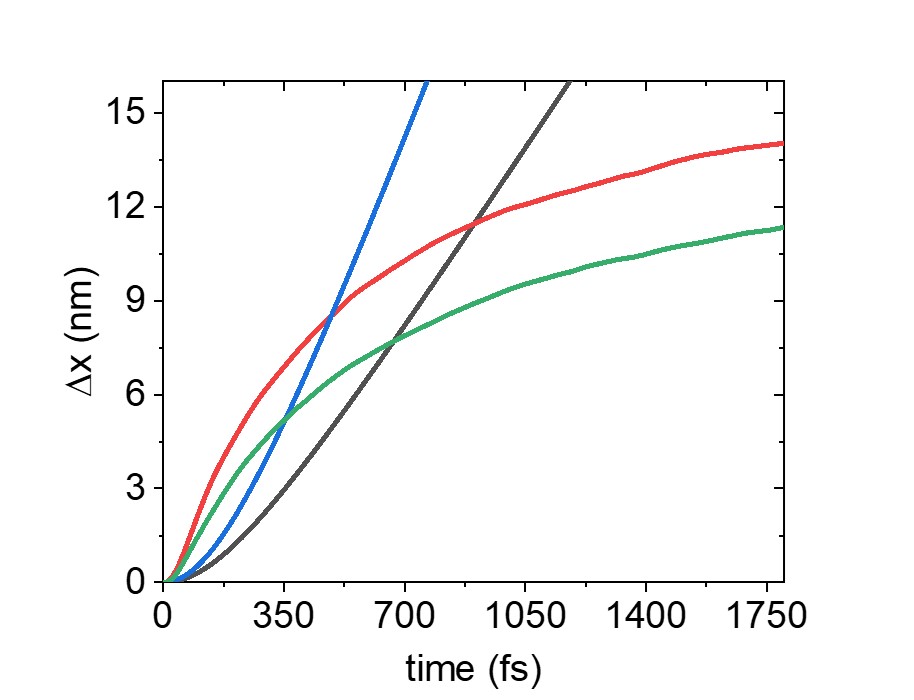}
\caption{The root mean square displacement (Eq. (6) in the main text) calculated from numerical simulations, shows for different shapes of initial wavepacket spread on an ordered lattice and on a lattice with static disorder. Black and red lines - the Gaussian wavepacket spread (initially takes the form of Eq. (2) in the main text) on an ordered and on a disordered lattice, respectively; blue and green lines – the $p$-like state wavepacket (initially takes the form of Eq. \eqref{eq:psi1}) spread on an ordered and on a disordered lattice, respectively. The width of the initial wavepacket is $D=20$nm. The hopping parameter is $J=0.5$eV and the disorder parameter is $\sigma=0.1eV$. }
\label{fig:S1}
\end{figure}

\begin{figure}[!htp]
\includegraphics[width=0.8\textwidth]{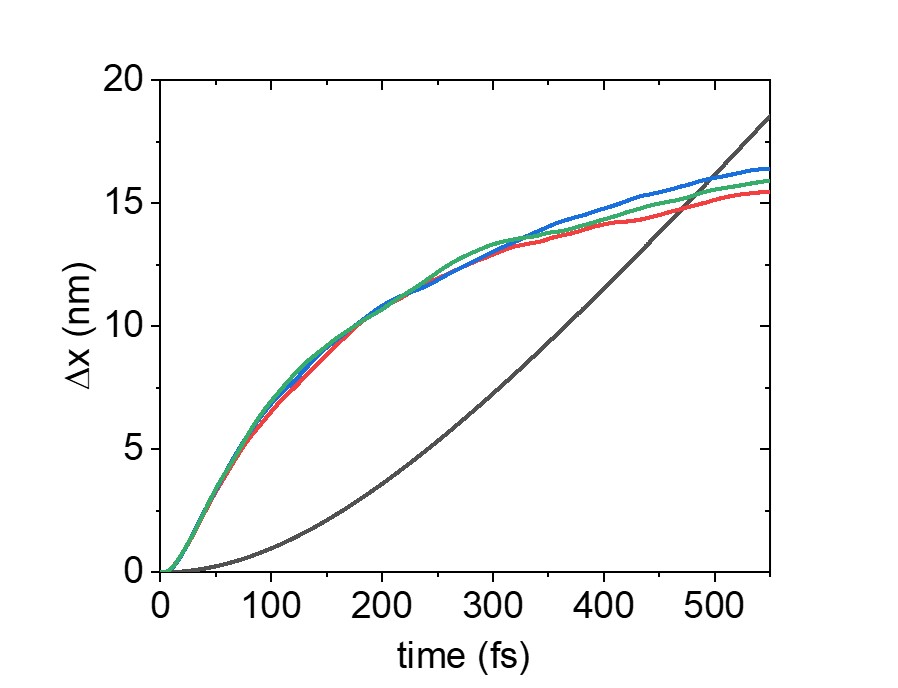}
\caption{The root mean square displacement (Eq. (6) in the main text) calculated from numerical simulations, shows for the Gaussian wavepacket spread on an ordered lattice and on a lattice with disorder, carrying different initial linear momentum, c.f. Eq. \eqref{eq:psi2}. Black line - ordered lattice (the result does not depend on the initial momentum); red line – static disorder for wavepacket with no initial momentum; blue line – static disorder for a moving wavepacket with initial momentum $p=0.5\hbar/D$; green line – static disorder for a moving wavepacket with initial momentum $p=\hbar/D$. The width of the initial wavepacket is $D=20$nm. The hopping parameter is $J=0.5$eV and the disorder parameter is $\sigma=0.1eV$. }
\label{fig:S2}
\end{figure}

In addition to the results shown in the main text, we provide here simulation results showing the effect of static disorder on the spread of exciton wavepackets of different shape and linear momentum. Figure \ref{fig:S1} shows the result obtained for $\Delta x(t)$ (Eq. (6) in the main text) for an initial wavepacket of the form 
\begin{equation}
\Psi(x,t=0)=\frac{\sum_n(na)e^{-\frac{(na)^2}{D^2}}\phi(x-na)}{\sqrt{\sum_n(na)^2e^{-2\left(\frac{na}{D}\right)^2}}},
\label{eq:psi1}
\end{equation}
i.e., a $p$-like state. Figure \ref{fig:S2} shows the similar result for the initial wavepacket
\begin{equation}
\Psi(x,t=0)=\frac{\sum_ne^{-\frac{(na)^2}{D^2}-ip(x-na)/\hbar}\phi(x-na)}{\sqrt{\sum_ne^{-2\left(\frac{na}{D}\right)^2}}},
\label{eq:psi2}
\end{equation}
that carries an initial linear momentum $p$. While Eq. \eqref{eq:psi1} is not of the type that will be excited by a short near-field pulse, these simulations serve to show the generality of our observations. In particular, note that the effect of an initial linear momentum (Fig. \ref{fig:S2}) is minimal.

\section{The short time evolution and the derivation of the auxiliary function $\tilde{\chi}(u;s)$}
\label{app.1}
\setcounter{equation}{2}
This section introduces the short time approximation and derives an expression Eq. \eqref{eq:chi} for an auxiliary function $\tilde{\chi}(u;s)$ (defined in Eq. \eqref{eq:chi0}) needed for a later evaluation of the short time evolution of the mean square displacement. Start from Eq. (12) in the main text and consider now the average that appears in the integrand on the RHS. Using the identity \cite{Frohlich,Ovchinnikov,Novikov}
\begin{equation}
\langle \xi(x)\Phi\{\xi\}\rangle_E=\int_{-\infty}^\infty dy\langle\xi(x)\xi(y)\rangle_E\left\langle\frac{\delta\Phi}{\delta\xi(y)}\right\rangle_E,
\end{equation}
which is valid for a Gaussian process $\xi(x)$ with zero mean and for any functional $\Phi\{\xi\}$, we have
\begin{align}
    \langle\tilde{\beta}(k_1,q)\delta(q'-k_2)\tilde{\rho}(q,q';t)\rangle_E=\int dxdy\langle\tilde{\beta}(k_1,q)\tilde{\beta}(x,y)\rangle_E\left\langle\frac{\delta\tilde{\rho}(q,q';t)}{\delta\tilde{\beta}(x,y)}\right\rangle_E\delta(q'-k_2),
    \label{eq:avefun0}
\end{align}
which gives 
\begin{align}
&\frac{\partial\langle \tilde{\rho}(k_1,k_2;t)\rangle_E}{\partial t}
=-\frac{i}{\hbar}(2J\cos(k_1)-2J\cos(k_2))\langle\tilde{\rho}(k_1,k_2;t)\rangle_E\notag\\
&-\frac{i}{2\pi\hbar}\int_{-\pi}^\pi dqdxdy\left(\langle\tilde{\beta}(k_1,q)\tilde{\beta}(x,y)\rangle_E\left\langle\frac{\delta\tilde{\rho}(q,k_2,t)}{\delta\beta(x,y)}\right\rangle_E-\langle\tilde{\beta}(q,k_2)\tilde{\beta}(x,y)\rangle\left\langle\frac{\delta\tilde{\rho}(k_1,q;t)}{\delta\beta(x,y)}\right\rangle_E\right).
\label{eq:rhoSF}
\end{align}
We note 
\begin{subequations}
\begin{align}
    \langle\tilde{\beta}(k_1,q)\tilde{\beta}(x,y)\rangle_E=g(0)\sum_me^{i(q+y-k_1-x)m},\\
    \langle\tilde{\beta}(q,k_2)\tilde{\beta}(x,y)\rangle_E=g(0)\sum_me^{i(y+k_2-q-x)m},
\end{align}
    \label{eq:beta}
\end{subequations}
and to identify the averaged derivatives of the RHS of Eq. \eqref{eq:rhoSF}, we take the functional derivative on both sides of Eq. (12) in the main text without ensemble average, which yields
\begin{align}
    \frac{\partial}{\partial t}\frac{\delta\tilde{\rho}(k_1,k_2;t)}{\delta\tilde{\beta}(x,y)}&=-\frac{i}{\hbar}(2J\cos(k_1)-2J\cos(k_2))\frac{\delta\tilde{\rho}(k_1,k_2,t)}{\delta\beta(x,y)}\notag\\
    &-\frac{i}{2\pi\hbar}\int_{-\pi}^\pi dqdq'\left[\delta(k_1-x)\delta(q-y)\delta(q'-k_2)-\delta(q'-x)\delta(k_2-y)\delta(q-k_1)\right]\tilde{\rho}(q,q';t)\notag\\
    &-\frac{i}{2\pi\hbar}\int_{-\pi}^\pi dqdq'[\tilde{\beta}(k_1,q)\delta(q'-k_2)-\tilde{\beta}(q'-k_2)\delta(q-k_2)]\frac{\delta\tilde{\rho}(q,q';t)}{\delta\tilde{\beta}(x,y)}.
    \label{eq:functrho}
\end{align}
For small absolute values of $\beta$ parameters, terms on the the last line maybe dropped. Further, assuming the same (vanishing at $t=0$) initial condition for any realization of $\beta$, is equivalent to a physical picture in which the disorder is switched on at $t=0$, it amounts to taking the functional derivative $\delta\tilde{\rho}/\delta\tilde\beta$ as zero at $t=0$. This gives us, from Eq. \eqref{eq:functrho}, 
\begin{tiny}
\begin{equation}
    \left\langle\frac{\delta\tilde{\rho}(k_1,k_2;t)}{\delta\tilde{\beta}(x,y)}\right\rangle_E=-\frac{i}{2\pi\hbar}\int_0^te^{-\frac{i2J(\cos(k_1)-\cos(k_2))}{\hbar}(t-\tau)}\int_{-\pi}^\pi dqdq'\left[\delta(k_1-x)\delta(q-y)\delta(q'-k_2)-\delta(q'-x)\delta(k_2-y)\delta(q-k_1)\right]\langle\tilde{\rho}(q,q';\tau)\rangle_Ed\tau.
    \label{eq:rhobetat}
\end{equation}
\end{tiny}
Substituting Eq. \eqref{eq:beta} and Eq. \eqref{eq:rhobetat} into Eq. \eqref{eq:rhoSF}, we obtain
\begin{align}
    \frac{\partial\langle \tilde{\rho}(k_1,k_2;t)\rangle_E}{\partial t}&=-\frac{i}{\hbar}(2J\cos(k_1)-2J\cos(k_2))\langle\tilde{\rho}(k_1,k_2;t)\rangle_E\notag\\
    &-\frac{g(0)}{2\pi\hbar^2}\int_{-\pi}^\pi dq\int_0^te^{-\frac{i2J(\cos q-\cos k_2)}{\hbar}(t-\tau)}\langle\tilde{\rho}(k_1,k_2;\tau)\rangle_Ed\tau\notag\\
    &-\frac{g(0)}{2\pi\hbar^2}\int_{-\pi}^\pi dq\int_0^te^{-\frac{i2J(\cos k_1-\cos q)}{\hbar}(t-\tau)}\langle\tilde{\rho}(k_1,k_2;\tau)\rangle_Ed\tau\notag\\
    &+\frac{g(0)}{2\pi\hbar^2}\int_{-\pi}^\pi dq\int_0^te^{-\frac{i2J(\cos q-\cos k_2)}{\hbar}(t-\tau)}\langle\tilde{\rho}(q,q+k_2-k_1;\tau)\rangle_Ed\tau\notag\\
        &+\frac{g(0)}{2\pi\hbar^2}\int_{-\pi}^\pi dq\int_0^te^{-\frac{i2J(\cos k_1-\cos q)}{\hbar}(t-\tau)}\langle\tilde{\rho}(q+k_1-k_2,q;\tau)\rangle_Ed\tau.
    \label{eq:exactrhot}
\end{align}
The short time approximation is now made by disregarding the oscillating phases in exponential in Eq. \eqref{eq:exactrhot}. This approximation is valid for $t<\hbar/J$ in our dimensionless unit $t<1$. This leads to
\begin{align}
    \frac{\partial\langle \tilde{\rho}(k_1,k_2;t)\rangle_E}{\partial t}&\approx-\frac{i}{\hbar}(2J\cos(k_1)-2J\cos(k_2))\langle\tilde{\rho}(k_1,k_2;t)\rangle_E\notag\\
    &-\frac{g(0)}{2\pi\hbar^2}\int_{-\pi}^\pi dq\int_0^t\langle\tilde{\rho}(k_1,k_2;\tau)\rangle_Ed\tau\notag\\
    &-\frac{g(0)}{2\pi\hbar^2}\int_{-\pi}^\pi dq\int_0^t\langle\tilde{\rho}(k_1,k_2;\tau)\rangle_Ed\tau\notag\\
    &+\frac{g(0)}{2\pi\hbar^2}\int_{-\pi}^\pi dq\int_0^t\langle\tilde{\rho}(q,q+k_2-k_1;\tau)\rangle_Ed\tau\notag\\
        &+\frac{g(0)}{2\pi\hbar^2}\int_{-\pi}^\pi dq\int_0^t\langle\tilde{\rho}(q+k_1-k_2,q;\tau)\rangle_Ed\tau.
    \label{eq:shortrhot}
\end{align}
Upon performing the Laplace transformation, $\mathcal{L}[f(t)]:=\hat{f}(s)=\int_0^\infty e^{-st}f(t)dt$, we have 
\begin{align}
    s\langle\tilde\rho(k_1,k_2;s)\rangle_E&=\tilde\rho_(k_1,k_2;t=0)-\frac{i2J(\cos k_1-\cos k_2)}{\hbar}\langle\tilde\rho(k_1,k_2;s)\rangle_E\notag\\
    &-\frac{2g(0)}{s\hbar^2}\langle\tilde\rho(k_1,k_2;s)\rangle_E\notag\\
    &+\frac{g(0)}{2\pi\hbar^2}\int_{-\pi}^\pi\left(\frac{\langle\tilde\rho(q,q+k_2-k_1;s)\rangle_E}{s}+\frac{\langle\tilde\rho(q+k_1-k_2,q;s)\rangle_E}{s}\right) dq.
\end{align}

%\begin{figure}[!tp]
%\includegraphics[width=0.48\textwidth]{Fig.S1}
%\caption{The time evolution of the deviation of the root-mean-square of the displacement from its initial position calculated from Eq. \eqref{eq:fit1} (solid lines), and from our numerical simulation (dots). Black line and dots - ordered case ($g(0)=0$). Red line and dots - disordered system ($g(0)=0.09J^2$). In both cases, the hopping parameter is $J=1$, the lattice spacing is $a=1$  and the initial exciton wavepacket width is $D=10$. The simulation cell contains $N=501$ lattice points. An average of $n=60$ realizations is taken and the estimated error in the numerical collection of the disordered system is smaller ($<10\%$) than the size of point. }
%\label{fig:fit1}
%\end{figure}

Using the fact that
\begin{equation}
    \int_{-\pi}^\pi dq\langle\tilde{\rho}'(q+k_1,q+k_2;\tau)\rangle_E
\end{equation}
depends on $k_1-k_2$ only, setting $u=k_1-k_2,p=(k_1+k_2)/2$ and denoting $\hat{R}(p,u;s)\equiv\tilde{\rho}(k_1,k_2;s)$, we have 

\begin{equation}
    \langle\hat{R}(p,u;s)\rangle_E=\frac{\hat{R}(p,u;t=0)+\frac{2g(0)}{\hbar^2s}\frac{1}{2\pi}\int_{-\pi}^\pi\langle\hat{R}(p+q,u;s)\rangle_E dq}{s-\frac{i4J\sin p\sin(\frac{u}{2})}{\hbar}+\frac{2g(0)}{\hbar^2s}}.
    \label{eq:fit1}
\end{equation}

We note that, when $g(0)=0$, then Eq. \eqref{eq:fit1} reduces to
\begin{equation}
    \langle\tilde{\rho}(p,u;s)\rangle_E=\frac{\rho(p,u;t=0)}{s-\frac{i}{\hbar}4J\sin(p)\sin\left(\frac{u}{2}\right)}.
\end{equation}
Using Eq. (13) in the main text, one can show $\langle\tilde{\rho}(p,u;s)\rangle_E\sim1/s^3$. In other words, we return to the ballistic ($\langle x^2\rangle\sim t^2$) motion in ordered lattice.

Note that $\int_{-\pi}^\pi\langle\hat{R}(q+p,u;s)\rangle_E dq$ does not depend on $p$. Defining the integrated density matrix
\begin{equation}
    \tilde{\chi}(u;s)\equiv\frac{1}{2\pi}\int_{-\pi}^\pi\langle\hat{R}(q,u;s)\rangle_E dq=\frac{1}{2\pi}\int_{-\pi}^\pi\langle\hat{R}(q+p,u;s)\rangle_E dq.
    \label{eq:chi0}
\end{equation}
We get an explicit expression for this function by integrating Eq. \eqref{eq:fit1} over the variable $p$:
\begin{equation}
    \tilde{\chi}(u;s)=I_1+\frac{2I_2g(0)\tilde{\chi}(u;s)}{\hbar^2},
    \label{eq:chi}
\end{equation}
which is equivalent to Eq. (15) where $I_1$ and $I_2$ are given by Eq. (16) in the main text.
The initial function $R(p,u;t=0)$ is obtained using the density matrix associated with the initial wavepacket function,
\begin{equation}
\rho_{m,n}(t=0)=\frac{e^{-\left(\frac{ma}{D}\right)^2}e^{-\left(\frac{na}{D}\right)^2}}{\sum_je^{-2\left(\frac{ja}{D}\right)^2}}.
\label{eq:rhomn0}
\end{equation}
The corresponding spatial Fourier transform is
\begin{align}
&\rho(p,u;t=0)=\sum_{m,n}\frac{e^{-\left(\frac{ma}{D}\right)^2}e^{-\left(\frac{na}{D}\right)^2}}{\sum_je^{-2\left(\frac{ja}{D}\right)^2}}e^{-im\left(p+\frac{u}{2}\right)}e^{in\left(p-\frac{u}{2}\right)}\notag\\
=&\frac{e^{-\frac{(D/a)^2}{4}\left(2p^2+\frac{u^2}{2}\right)}}{\sum_je^{-2\left(\frac{j}{(D/a)^2}\right)^2}}\sum_{m}e^{-\frac{1}{(D/a)^2}\left(m+\frac{i(D/a)^2p}{2}+\frac{i(D/a)^2u}{4}\right)^2}\notag\\
&\quad\cdot\sum_{n}e^{-\frac{1}{(D/a)^2}\left(n+\frac{i(D/a)^2p}{2}-\frac{i(D/a)^2u}{4}\right)^2}\notag\\
&=\sqrt{2\pi (D/a)^2}e^{-\frac{(D/a)^2}{4}\left(2p^2+\frac{u^2}{2}\right)},
\label{eq:Rt0}
\end{align}
where we have converted the discrete sum to continuous integral, e.g. $\sum_{j}\rightarrow\int dj$.

\section{Equivalence of analytical and numerical approaches for ordered lattices}
\label{app.2}
\setcounter{equation}{18}
Here we express the mean square displacement in terms of the function $\tilde{\chi}(u;s)$ which is defined by  Eq. \eqref{eq:chi0} and show that for ordered lattices, analytical and numerical calculations are equivalent.
On one hand, we note 
\begin{align}
    \tilde{\chi}(u;s)&=\frac{1}{2\pi}\int_{-\pi}^\pi\tilde{\rho}(q,u;s)dq\notag\\
    &=\frac{1}{2\pi}\int_{-\pi}^\pi \sum_{m,n}\rho_{m,n}(s)e^{-im\left(p+\frac{u}{2}\right)}e^{in\left(p-\frac{u}{2}\right)}dp\notag\\
    &=\sum_{m,n}\delta_{m,n}\rho_{m,n}(s)e^{-i(m+n)\frac{u}{2}}\notag\\
    &=\sum_m\rho_{m,m}(s)e^{-imu}.
\end{align}
In $t$-space, it is $\tilde{\chi}(u;t)=\sum_m\rho_{m,m}(t)e^{-imu}$, which leads Eq. (13) in the main text.
On the other hand, recall the formula of mean square displacement in numerical simulation: 
\begin{widetext}
\begin{align}
\langle x^2(t)\rangle=\langle\Psi(t)|\hat{x}^2|\Psi(t)\rangle=\frac{1}{N^2\sum_me^{-2\left(\frac{ma}{D}\right)^2}}\sum_{j=-(N-1)/2}^{(N-1)/2}\left|\sum_{k_1}\sum_me^{-\left(\frac{ma}{D}\right)^2}e^{ik_1ma} e^{-ik_1ja}e^{iE_{k_1}t/\hbar}(ja)\right|^2.
\end{align}
\end{widetext}
where for convenience, we let $N$ be odd.
When $N\rightarrow\infty$, we have $\sum_m\rightarrow\int dm$ and $\sum_k/N\rightarrow \int_{-\pi}^\pi dk/{2\pi}$. In $s$-space or regular time space, using Eqs. (\ref{eq:rhomn0}) and (\ref{eq:Rt0}), we have
\begin{align}
&\langle x^2(s)\rangle=\sum_jj^2\frac{1}{2\pi}\int_{-\pi}^\pi dk_1 e^{ijk_1}\frac{1}{2\pi}\int_{-\pi}^\pi dk_2 e^{-ijk_2}\frac{\tilde{\rho}(k_1,k_2;t=0)}{s+\frac{i}{\hbar}(o_{k_1}-o_{k_2})}\notag\\
&\langle x^2(t)\rangle=\sum_jj^2\frac{1}{2\pi}\int_{-\pi}^\pi dk_1 e^{ijk_1}\frac{1}{2\pi}\int_{-\pi}^\pi dk_2 e^{-ijk_2}\tilde{\rho}(k_1,k_2;t)\notag\\
&=\sum_jj^2\frac{1}{2\pi}\int_{-\pi}^\pi dk_1 e^{ijk_1}\frac{1}{2\pi}\int_{-\pi}^\pi dk_2 e^{-ijk_2}\sum_{m,n}\rho(m,n;t) e^{-ik_1m+ik_2n}\notag\\
&=\sum_jj^2\frac{1}{2\pi}\int_{-\pi}^\pi dk_1 e^{i(j-m)k_1}\frac{1}{2\pi}\int_{-\pi}^\pi dk_2 e^{i(n-j)k_2}\sum_{m,n}\rho(m,n;t)\notag\\
&=\sum_jj^2\delta_{j,m}\delta_{j,n}\rho(m,n;t)\notag\\
&=\sum_jj^2\rho(j,j;t).
\label{ap:TBGa}
\end{align}
Thus, we verify the equivalence between numerical and analytical approaches.

\newpage

\bibliography{Anderson}